# Infrared Compositional Measurements in Comet C/2017 K2 (Pan-STARRS) at Heliocentric Distances Beyond 2.3 AU

Chemeda Ejeta[1], Erika Gibb[1], Michael A. DiSanti [2], Hideyo Kawakita [3], Boncho P. Bonev[4], Neil Dello Russo [5], Nathan Roth[2,6], Younas Khan[7], Adam J. McKay [8], Michael R. Combi[9], Lori Feaga[10], Mohammad Saki [1], Ronald J. Vervack Jr.[5], Yinsi Shou[9]

[1] Department of Mathematics, Physics, Astronomy, and Statistics, University of Missouri-St. Louis, Saint Louis, MO 63121, USA

[2] Solar System Exploration Division, Planetary Systems Laboratory, MS 693, NASA Goddard Space Flight Center, Greenbelt, MD 20771, USA

[3] National Astronomical Observatory of Japan, 2-21-1 Osawa, Mitaka, Tokyo 181-8588, Japan

[4] Department of Physics, American University, Washington, DC 20016, USA

[5] Space Department, Johns Hopkins University Applied Physics Laboratory, Laurel, MD 20723, USA

[6] Catholic University of America, Washington, DC

[7] University of Alabama at Birmingham, Birmingham, AL, USA

[8] Department of Physics and Astronomy, Appalachian State University, Boone, NC 28608-2106

[9] Department of Climate and Space Sciences and Engineering, University of Michigan, Ann Arbor, MI 48109, USA

[10] University of Maryland, Department of Astronomy, College Park, MD 20742

# Abstract

Comet C/2017 K2 (Pan-STARRS) provided a rare opportunity to investigate the evolution of coma composition and outgassing patterns over a transitional heliocentric distance ($R_h$) range where activity drivers in comets are thought to change from "hypervolatile" (CO, $CH_4$, $C_2H_6$, and/or $CO_2$)-dominated to $H_2O$-dominated. We performed high-resolution ($\lambda/\Delta\lambda \approx 25{,}000 - 42{,}000$), cross-dispersed, near-infrared spectroscopy of C/2017 K2 with iSHELL at the NASA Infrared Telescope Facility (IRTF) and NIRSPEC at the W. M. Keck 2 Observatory. We report gas rotational temperatures ($T_{rot}$) and molecular production rates (Q; mol/s) or upper limits for the "hypervolatile" species $CH_4$, CO, and $C_2H_6$, together with less volatile ices ($CH_3OH$, $H_2O$, HCN, $C_2H_2$, $NH_3$, and OCS) over a range of pre-perihelion distances, $R_h$= 3.15 - 2.35 au. We also report (or stringently constrain) abundance ratios (mixing ratios) of the targeted species with respect to CO, $C_2H_6$, and (when detected) $H_2O$. All volatiles were enriched relative to water in C/2017 K2 when compared to their mean values among Oort Cloud comets, whereas abundances relative to $C_2H_6$ were consistent with their average values from other long-period comets.

# 1. INTRODUCTION

Because of their ~ 4.5 Gyr residence in the far-out, cold Oort cloud region, long-period comets are thought to be least affected by solar radiation, making them the most primitive small icy bodies in the solar system. They can, therefore, be used to constrain conditions present in the protoplanetary disk when (and where) they formed and, thus, the subsequent evolution of the early solar system (Dello Russo et al. 2016). Although coma volatile abundances relative to $H_2O$ (the most abundant ice in most comets) provide an excellent proxy for bulk nucleus composition when comets are close to the Sun, more distant comets may be driven by preferential outgassing of hypervolatile species, such as CO and $CO_2$ (A'Hearn et al. 2012; Ootsubo et al. 2012) or alternatively exothermic crystallization of amorphous $H_2O$ ice (e.g., Jewitt 2009; Guilbert-Lepoutre et al. 2012; Prialnik & Bar-Nun, 1992). For comets that have been observed to be active at large distances post-perihelion, heat acquired near perihelion could be slowly transferred into the interior of the nucleus through conduction. This could cause subsurface volatiles to activate later and fuel distant outbursts (e.g., Prialnik, 1992).

The critical region for the transition to water-driven sublimation for an inbound comet is thought to be between ~2.5-3 au (e.g., Ootsubo et al. 2012, Meech and Svoren 2004), which is supported by Biver et al. (2002) who conducted an investigation of the outgassing behavior of various molecular species (including OH, CO, HCN, $CH_3OH$, and $H_2CO$) at radio wavelengths for comet Hale-Bopp spanning a wide heliocentric distance ($R_h$) range of 0.9-14 au and found that the OH production rate (a proxy for $H_2O$ production rate) surpassed that of CO within about 3 au. Thus, to more accurately determine the volatile abundances of comets for comparison to protoplanetary disk models, it is crucial to understand when volatile species in comets are fully activated.

The discovery of long-period comet C/2017 K2 (hereafter K2), by the Panoramic Survey Telescope and Rapid Response System (Pan-STARRS) on UT 2017 May 21 at a heliocentric distance $R_h$ = 15.9 au (Wainscoat et al. 2017), provided a rare opportunity for a detailed study of the evolution of cometary activity at larger $R_h$. Analysis of prediscovery archival data later showed that K2 was active even at $R_h$ = 23.7 au (Jewitt et al. 2017; Meech et al. 2017; Hui et al. 2018). As this was well beyond the distance at which water crystallization should occur, the activity was likely due to the sublimation of hypervolatile ices (Meech et al. 2017; Jewitt et al. 2017), which is supported by the detection of CO in K2 at $R_h$ = 6.72 au (Yang et al., 2021). Atomic oxygen emission at $R_h$ = 2.8 au also suggested significant $CO_2$-driven activity (Cambianica et al. 2023).

In this paper, we report high-resolution, near-infrared spectroscopic observations of K2 on nine distinct dates spanning $R_h$ = 3.15 - 2.35 au pre-perihelion with the goal of investigating the distance at which water and other less volatile species begin sublimating. In section 2, we discuss observations of K2. In section 3, we present data reduction and analysis. In section 4, we present a detailed discussion, and in section 5, we summarize our observational results.

## 2. OBSERVATIONS

Observations of K2 were carried out on nine dates from UT 2022 May 18-Aug 21, spanning $R_h$ ~ 3.15 – 2.35 au pre-perihelion. Comet K2 had a perihelion distance of ~ 1.80 au on UT 2022 December 20. The observations were performed using both iSHELL (Rayner et al. 2022) at the NASA Infrared Telescope Facility (IRTF) and NIRSPEC (Martin et al. 2016, 2018) at Keck II, to sample (primarily) the "hypervolatile" species CO, $CH_4$, and $C_2H_6$, and to test the extent to which $H_2O$ was activated in K2 over this range of $R_h$.

The observations were acquired using a 0.75" wide x 15" long slit for iSHELL and a 0.43" x 24" slit for NIRSPEC, resulting in a resolving power ($\lambda/\Delta\lambda$) ~ 4.2 × $10^4$ and ~ 2.5 × $10^4$, respectively. The comet frames were obtained using an exposure sequence A-B-B-A, with A and B beams placed equidistant to either side of the slit center and separated by half its length (e.g., DiSanti et al. 2006; Bonev et al. 2021). The arithmetic operation A – B – B + A isolates the net comet signal within each 2048x2048-pixel frame (e.g., see Figure 1 of DiSanti et al. 2017).

We acquired spectra using three different iSHELL settings: Lp1 ranging from ~3.3 – 3.6 μm to sample $C_2H_6$, $CH_4$, $CH_3OH$, and OH* (or prompt OH emission), M2 covering ~4.5 -5.2 μm to sample CO, OCS, and $H_2O$, and a customized L-band setting (L-custom) spanning ~2.8-3.1 μm, to sample $H_2O$, HCN, $C_2H_2$, and $NH_3$. The NIRSPEC observations were performed using the KL1 setting to sample $H_2O$, $C_2H_6$, and $CH_3OH$ and the KL2 setting to sample $H_2O$, HCN, $C_2H_2$, $CH_4$, $C_2H_6$, and $CH_3OH$ (see Table 1). CO and $CH_4$ have the lowest vacuum sublimation temperatures of any cometary ices targeted using IR spectroscopy. Both require sufficiently large geocentric velocities (d$\Delta$/dt), to Doppler-shift cometary emission lines from the (opaque) cores of their strong terrestrial counterpart absorptions and into regions of higher atmospheric transmittance. Table 1 includes dates on which d$\Delta$/dt ~ 12 km s$^{-1}$ or higher, thereby enabling measurement of these two molecules species. Together with only slightly less volatile $C_2H_6$, we refer to CO and $CH_4$ as "hypervolatiles."

**Table 1**
Observing Log for K2

| 2022 UT Date | Setting[a] | Target | UT time | $R_h$ (au) | $\Delta$ (au) | d$\Delta$/dt (km s$^{-1}$) | $T_{int}$ (minutes) | Slit PA (°) |
|---|---|---|---|---|---|---|---|---|
| May 18 | Lp1 | BS 6117 | 11:16–11:28 | --- | --- | --- | --- | --- |
|  | M2 | BS 6117 | 11:32–11:42 | --- | --- | --- | --- | --- |
|  |  | K2 | 11:52–13:20 | 3.15 | 2.44 | -34.7 | 80 | 236 |
|  | Lp1 | K2 | 13:29–15:00 |  |  |  | 80 |  |
| June 08 | Lp1 | BS 5867 | 08:47 – 08:55 | --- | -- | -- | --- | --- |

| Date | Settinga | Source | UT (hh:mm) | $R_h$ (AU) | $\Delta$ (AU) | $d\Delta/dt$ (km s⁻¹) | $T_{int}$ (min) | PA (°) |
|---|---|---|---|---|---|---|---|---|
| | M2 | BS 5867 | 09:03- 09:13 | --- | --- | --- | --- | --- |
| | K2 | 09:23- 11:27 | 2.97 | 2.08 | -25.4 | 96 | 207 | |
| | Lp1 | K2 | 11:36-12:57 | | | | 72 | |
| June 18 | Lp1 | BS 5867 | 08:18- 08:27 | --- | --- | --- | --- | --- |
| | M2 | BS 5867 | 08:35-08:47 | --- | --- | --- | --- | --- |
| | | K2 | 08:57- 10:47 | 2.88 | 1.95 | -19.0 | 84 | 183 |
| | Lp1 | K2 | 10:55-12:25 | | | | 80 | |
| Jul 28* | KL1 | K2 | 06:38-08:03 | 2.54 | 1.85 | +8.9 | 56 | 107 |
| | KL1 | BS6175 | | | | | | |
| | KL2 | K2 | 08:04-09:29 | 2.54 | 1.85 | +9.1 | 60 | |
| | KL2 | BS7950 | | | | | | |
| Aug 04* | KL2 | K2 | 08:04-08:49 | 2.48 | 1.89 | +12.3 | 28 | 104 |
| | KL2 | BS | | | | | | |
| Aug 10 | Lp1 | K2 | 05:07-06:35 | 2.43 | 1.93 | 14.2 | 64 | 102 |
| | M2 | K2 | 06:46-08:45 | | | | 90 | |
| | M2 | BS 7236 | 08:58 -09:07 | --- | --- | --- | --- | --- |
| | Lp1 | BS 7236 | 09:09-09:17 | --- | --- | --- | --- | --- |
| Aug 19 | Lp1 | K2 | 05:11-07:26 | 2.36 | 2.01 | 16.6 | 120 | 101 |
| | | BS 7236 | 07:36 – 07:47 | --- | --- | --- | --- | --- |
| Aug 20 | L-custom | K2 | 05:09 – 07:28 | 2.35 | 2.02 | 16.8 | 116 | 101 |
| | | BS 7236 | 07:39 – 07:50 | --- | --- | --- | --- | --- |
| Aug 21 | M2 | K2 | 05:04 – 07:32 | 2.35 | 2.03 | 16.9 | 106 | 101 |
| | | BS 7236 | 07:42 – 07:54 | --- | --- | --- | --- | --- |

**Notes.** $R_h$, $\Delta$, and $d\Delta/dt$ are the heliocentric distance, geocentric distance, and geocentric velocity, respectively of comet K2, and $T_{int}$ is the total on-source integration time. The slit position angle (PA) was oriented along the projected Sun-comet line on all dates; those for NIRSPEC observations are marked '*'; iSHELL was used on all other dates.

[a] Spectral setting used to sample molecular lines.

## 3. DATA REDUCTION AND ANALYSIS

### 3.1 Initial Processing

We followed a data reduction procedure tailored to iSHELL and NIRSPEC spectra (e.g., DiSanti et al. 2017, 2021; Dello Russo et al. 2020; Bonev et al. 2021, 2023). Each echelle order was flat-fielded, dark subtracted, cleaned for cosmic ray hits, and spectrally and spatially straightened such that each row corresponded to a unique position along the slit and each column to a unique wavelength. For iSHELL (NIRSPEC) observations, comet spectra were extracted from the processed orders by summing signal over 15 (9) rows centered on the nucleus, defined as the position of peak emission intensity. We refer to these as "nucleus-centered" spectral extracts (nucleus-centered spectra), from which nucleus-centered production rates (see §3.3) were calculated. The nucleus-centered aperture consisted of a central square box (defined by the slit width), plus equivalent boxes to either side of the nucleus, hence a 1x3 rectangular nucleus-centered aperture.

Contributions from continuum and gaseous emissions (e.g., as shown in Fig. 1) were determined as outlined in DiSanti et al. (2017) and Bonev et al. (2021). Column burdens of absorbing species in the telluric atmosphere were retrieved using the NASA Planetary Spectrum Generator (Villanueva et al. 2018), by fitting a synthetic spectrum of atmospheric transmittance to IR flux standard star spectra obtained in each instrument setting using the widest available slit (4.0″ for iSHELL and 0.72″ for NIRSPEC). These column burdens were then applied to the comet spectra: the fully resolved ($\lambda/\Delta\lambda$ ~$10^6$) synthetic transmittance function was convolved to the resolving power of the comet data and scaled to the level of the comet continuum. The cometary emission lines were isolated by subtracting this modeled continuum.

Finally, each modeled g-factor (i.e., modeled line intensity) was corrected for the monochromatic atmospheric transmittance at its Doppler-shifted wavelength, based on the geocentric velocity of the comet at the time of the observation. Synthetic emission models for each targeted species were then compared to the observed line intensities. The g-factors used in this study were generated for $H_2O$ (Villanueva et al. 2012a), OH* (Bonev et al. 2006), HCN and $NH_3$ (Villanueva et al. 2013), $C_2H_2$ (see appendix C of Villanueva et al. 2011a,), $C_2H_6$ (Villanueva et al. 2011a), $CH_3OH$ (Villanueva et al. 2012b; DiSanti et al. 2013; Bonev et al., in prep), $CH_4$ (Gibb et al. 2003), and CO (DiSanti et al. 2001).

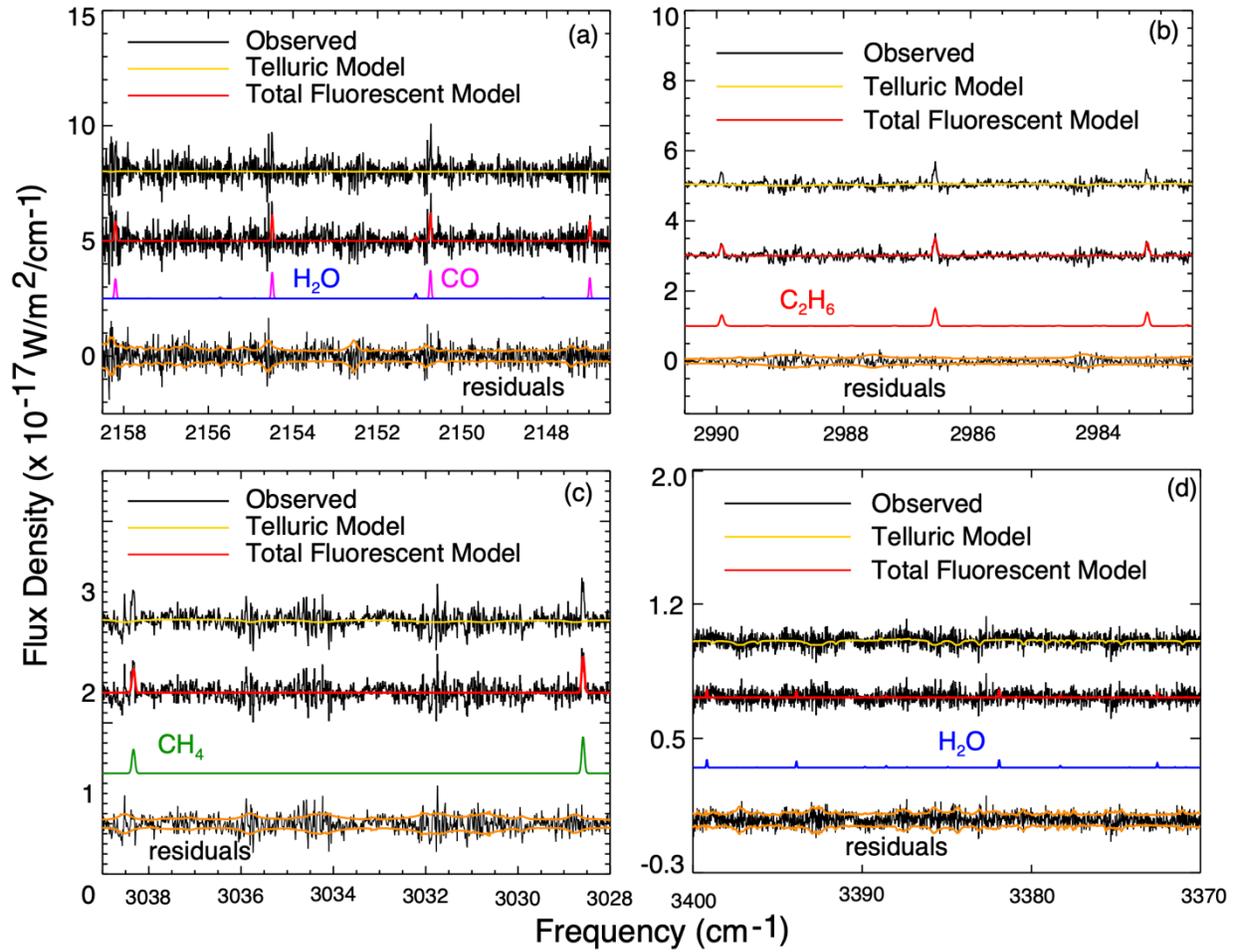

Figure 1. Sample spectra and best-fit fluorescent model of K2 on UT 2022 August 10 (a), August 19 [(b) and (c)], and August 20 (d). The observed spectra (black) are superimposed with the telluric model (yellow). Continuum-subtracted spectra with individual fluorescence emission models (color-coded by species) are shown below the observed spectrum in each panel. The bottom trace indicates the residual spectrum (after subtracting the modeled continuum and total molecular contributions), with the superimposed $\pm 1\sigma$ stochastic noise envelope shown in orange.

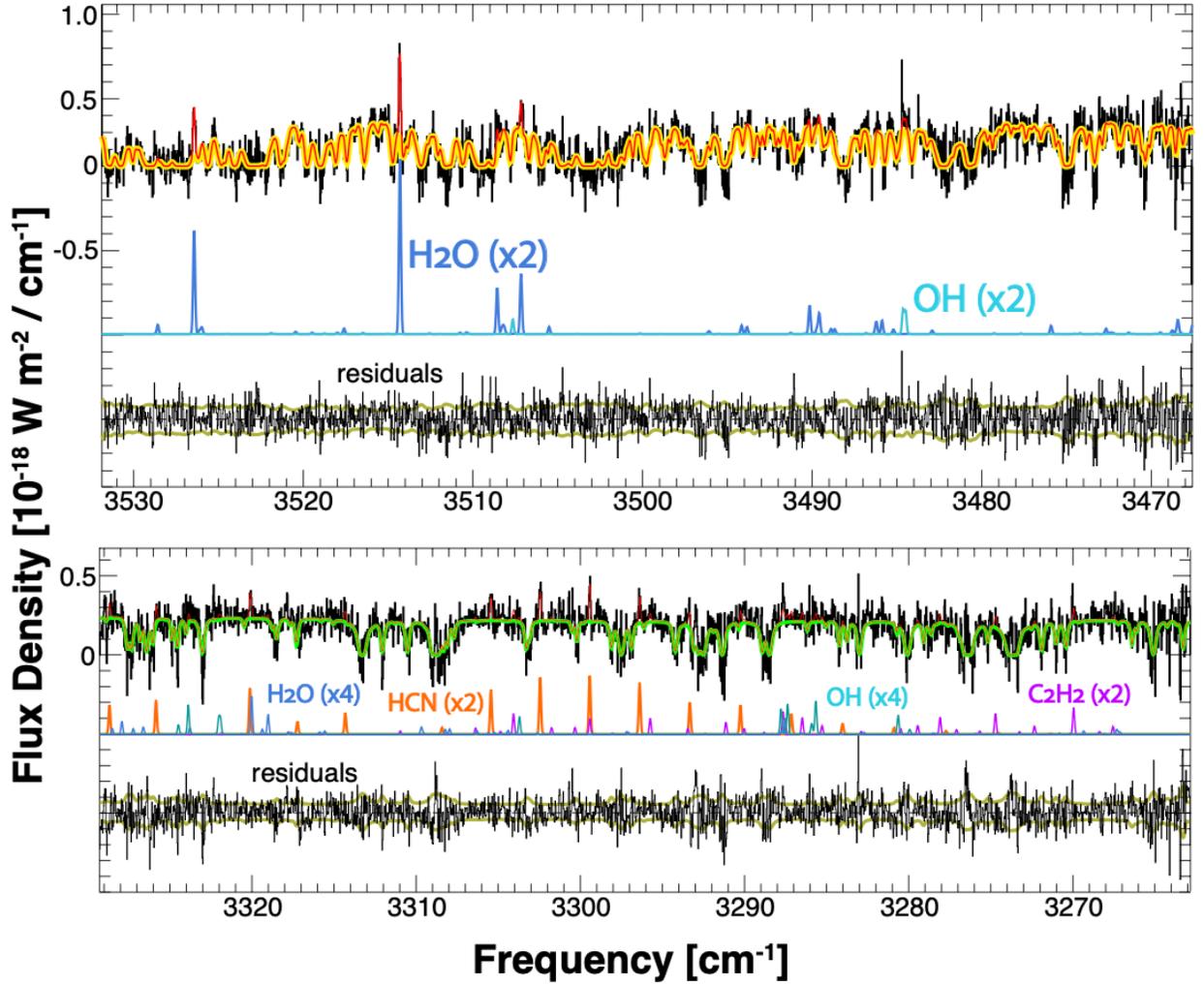

Figure 2. Sample spectra of $H_2O$, $OH^*$, HCN, and $C_2H_2$ acquired using the NIRSPEC KL1 (top panel) and KL2 (bottom panel) settings on UT 2022 July 28 with 56 and 60 minutes on the source, respectively (Table 1). The observed spectra are in black, with the telluric model overplotted in yellow (for KL1) and green (for KL2). The individual fluorescence models are indicated in different colors, with red showing the total fluorescence and telluric models. The residual spectra are displayed at the bottom of each panel with 1σ stochastic noise envelope superimposed.

### 3.2 Determination of Rotational Temperatures

The excitation conditions in cometary comae can be quantified through the rotational temperature ($T_{rot}$). Obtaining $T_{rot}$ involves analysis of $F_{line}/g$ vs. upper state rotational energy ($E_u$), for emission lines spanning a sufficient range of $E_u$, where $F_{line}$ is the transmittance-corrected molecular line flux, and g is the fluorescence efficiency (or g-factor) of observed ro-vibrational lines (see Eq. 1 and related discussion). At the optimum $T_{rot}$, the quantity $F_{line}/g$ is constant (within uncertainty) for all $E_u$; this $T_{rot}$ corresponds to the zero-slope best fit of $F_{line}/g$ vs. $E_u$ (Dello Russo et al. 2004; DiSanti et al. 2006).

For each date, we obtained consistent rotational temperatures for our nucleus centered apertures (see § 3.3), for all species having sufficient signal-to-noise ratio (SNR; see Table 2). When the SNR for a given species was insufficient, we assumed a $T_{rot}$ determined from species measured on the same night. This assumption is reasonable given that $T_{rot}$ obtained from contemporaneously measured volatile species (both within the same setting and across settings) are generally found to be consistent (e.g., see § 3.1 in Gibb et al. 2012), and thermodynamic models for the inner (collisional) coma also show roughly the same effective temperature for all species observed within a given field of view (Combi et al. 2004; Fougere et al. 2012). In particular, for the UT July 28 NIRSPEC data, production rates are listed for both 40 K and 50 K temperatures (see Table 2), as these are representative of those measured for $H_2O$, HCN, $C_2H_6$, and $CH_3OH$ on this date. We note that for each of these species, the production rates at 40 and 50 K agree within their 1-sigma uncertainties.

### 3.3. Molecular Production Rates and Mixing Ratios

The nucleus-centered molecular production rate, $Q_{nc}$ (molecules s$^{-1}$), is given as (DiSanti et al. 2014):

$$Q_{nc} = \frac{4\pi \Delta^2 F_i}{f(x)\tau_1 g_1}, \tag{1}$$

where $F_i$ (Wm$^{-2}$) is the flux of emission line "*i*" contained in a 1x3 rectangular aperture centered on the peak emission intensity, and corrected for telluric transmittance at its Doppler-shifted line frequency, $g_1$(W molecule$^{-1}$) and $\tau_1$(s) are the fluorescence efficiency (g-factor) and photo-dissociation lifetime, respectively at $R_h$ = 1 au (both are $R_h$-dependent through the relations $g = g_1 R_h^{-2}$ and $\tau = \tau_1 R_h^2$), *f(x)* is the fraction of molecules contained in the nucleus-centered beam assuming entirely native release (i.e., release directly from the nucleus), and Δ (m) is the geocentric distance.

The technique of obtaining production rates is based on the Q-curve analysis (e.g., Bonev 2005). The square aperture (as defined in § 3.1) was stepped along the spatial profile of each measured emission (SNR-permitting), and Q was computed at each step, assuming uniform symmetric outflow. A symmetrized Q-curve was produced by averaging the production rates at diametrically opposite positions about the nucleus, to average over potential asymmetric outflow and thereby obtain the most reliable production rate (Xie and Mumma 1996; also, see § 2.5 of Villanueva et al. 2011b) and thereby obtain the most reliable production rate. These symmetrized production rates invariably increase with distance from the nucleus, then level off (within uncertainty) to a terminal (i.e., total) production rate, $Q_{term}$. For species released directly from the nucleus (this is termed native release), the leveling-off distance depends on atmospheric seeing, which redistributes flux from the innermost coma into the wings of the profile. The increase of production rate from the nucleus-centered position ($Q_{nc}$) to the terminal (off-nucleus) region (defining $Q_{term}$) establishes a "growth factor" (GF ≡ $Q_{term}/Q_{nc}$), which accounts for the "lost" (i.e., redistributed) flux (this phenomenon is also referred to as slit loss). Generally, the same atmospheric seeing conditions, as well as positioning in the slit affects molecules similarly within the same instrument setting (Villanueva et al. 2011b), assuming purely native release. Therefore, when low SNR precluded determining a reliable GF, we assumed the value determined from one or more simultaneously (or, if not, contemporaneously; see §4.2) measured species. Relevant production rates and growth factors are indicated in Table 2.

The chemical taxonomy of comets is based on the chemical abundances of molecular species, expressed as abundance ratios (i.e., mixing ratios), traditionally defined as the ratio of the production rate of molecule 'X' to that of simultaneously or contemporaneously measured $H_2O$ [i.e., $Q(X)/Q(H_2O)$], usually the most abundant molecule in a cometary comae. Table 2 also lists mixing ratios with respect to simultaneously or contemporaneously measured CO and $C_2H_6$; these hypervolatiles serve as useful compositional baselines for comets at larger $R_h$, particularly when $H_2O$ is not securely detected (or when its sublimation is not fully activated). Even at smaller $R_h$, where water production is fully activated, the value alterative baselines for expressing mixing ratios has been utilized (e.g., Bonev et al. 2021, DiSanti et al. 2021).

**Table 2**
Volatile Composition of Comet K2 obtained using iSHELL and NIRSPEC

| Setting | Molecule | $T_{rot}^{(a)}$ (K) | $Q_{nc}^{(b)}$ ($10^{26}$ mol s$^{-1}$) | $GF_{measd/assumd}^{(c)}$ [$GF_{used}$] | $Q^{(d)}$ ($10^{26}$ mol s$^{-1}$) | $(Q_X/Q_{H2O})^{(e)}$ (%) | $[Q(X)/Q(C_2H_6)]^{(f)}$ | $(Q_X/Q_{CO})^{(g)}$ (%) |
|---|---|---|---|---|---|---|---|---|
| **2022 May 18, $R_h$ = 3.15 au, $\Delta$ = 2.44 au, $d\Delta/dt$ = -34.73 km s$^{-1}$** | | | | | | | | |
| M2 | CO | $18_{-4}^{+6}$ | 13.9 ± 1.7 | 2.40 ± 0.20 | 33.4 ± 5.1 | > 7.62 | 8.10 ± 1.67 | 100 |
| | | (20) | 14.2 ± 1.7 | | 34.1 ± 5.1 | > 7.37 | 8.08 ± 1.65 | 100 |
| | $H_2O$ | (18) | < 183 | [2.40 ± 0.20] | < 438 | | < 106 | < 1313 |
| | | (20) | < 193 | | < 463 | 100 | < 109 | < 1357 |
| Lp1 | $C_2H_6$ | $20_{-4}^{+7}$ | 1.69 ± 0.22 | | 4.25 ± 0.58 | > 0.92 | 1.00 | 12.5 ± 2.5 |
| | | (18) | 1.64 ± 0.22 | [2.52 ± 0.14] | 4.13 ± 0.57 | > 0.94 | 1.00 | 12.4 ± 2.5 |
| | $CH_4$ | (20) | 2.82 ± 0.39 | | 7.10 ± 1.01 | > 1.53 | 1.67 ± 0.32 | 20.8 ± 4.3 |
| | | (18) | 2.67 ± 0.43 | 2.52 ± 0.14 | 6.72 ± 1.10 | > 1.53 | 1.63 ± 0.34 | 20.1 ± 4.5 |
| **2022 June 08, $R_h$ = 2.97 au, $\Delta$ = 2.08 au, $d\Delta/dt$ = -25.37 km s$^{-1}$** | | | | | | | | |
| M2 | CO | $24_{-5}^{+7}$ | 15.2 ± 1.3 | 2.09 ± 0.10 | 31.7 ± 2.8 | >13.8 | 11.1 ± 1.8 | 100 |
| | | (28) | 16.0 ± 1.4 | | 33.4 ± 3.1 | > 13.6 | 11.1 ± 1.7 | 100 |
| | $H_2O$ | (24) | < 110 | [2.09±0.10] | < 231 | | < 81.1 | < 727 |
| | | (28) | < 117 | | < 246 | | < 81.5 | < 735 |
| Lp1 | $C_2H_6$ | (28) | 1.44 ± 0.18 | [2.09 ±0.10] | 3.01 ± 0.38 | > 1.23 | 1.00 | 9.02 ± 1.41 |
| | | (24) | 1.36 ± 0.18 | | 2.85 ± 0.37 | > 1.23 | 1.00 | 8.97 ± 1.43 |
| | CH$_4$ | $28^{+4}_{-3}$ | 4.06 ± 0.17 | | 8.49 ± 0.39 | > 3.46 | 2.82 ± 0.37 | 25.4 ± 2.6 |
| | | (24) | 3.62 ± 0.21 | | 7.59 ± 0.46 | > 3.29 | 2.67 ± 0.38 | 23.9 ± 2.6 |
| **2022 June 18, R$_h$ = 2.88 au, Δ = 1.95 au, dΔ/dt = -18.95 km s$^{-1}$** | | | | | | | | |
| M2 | CO | $34^{+9}_{-7}$ | 15.9 ± 1.3 | 2.54 ± 0.15 | 40.4 ± 8.5 | >12.0 | 8.86 ± 2.22 | 100 |
| | | (32) | 15.7 ± 1.3 | | 39.8 ± 8.38 | > 11.9 | 8.87 ± 2.22 | 100 |
| | H$_2$O | (34) | < 132 | [2.54± 0.15] | < 336 | | < 73.6 | < 831 |
| | | (32) | < 132 | | < 334 | | < 74.4 | < 839 |
| Lp1 | C$_2$H$_6$ | $32^{+5}_{-4}$ | 2.08 ± 0.28 | 2.16 ± 0.21 | 4.49± 0.61 | > 1.34 | 1.00 | 11.3 ± 2.8 |
| | | (34) | 2.12 ± 0.28 | [2.14± 0.15] | 4.56± 0.62 | > 1.36 | 1.00 | 11.3 ± 2.8 |
| | CH$_4$ | (32) | 4.01 ± 0.55 | 2.12 ± 0.21 | 8.48 ± 1.17 | > 2.54 | 1.92 ± 0.37 | 21.3 ± 5.4 |
| | | (34) | 4.28 ± 0.58 | | 9.06 ± 1.25 | > 2.70 | | 22.42± 5.66 |
| **2022 July 28**, R$_h$ = 2.54 au, Δ = 1.85 au, dΔ/dt = +8.88 km s$^{-1}$ (KL1), +9.06 km s$^{-1}$ (KL2)** | | | | | | | | |
| KL1 | H$_2$O | (40) | 69.3 ± 4.7 | 1.78 ± 0.18 | 125.4 ± 10.7 | 100 | 33.46 ± 2.65 | 307 ± 67 |
| | | (50) | 69.5 ± 5.1 | | 128.1 ± 11.2 | 100 | 30.74 ± 2.77 | 313 ± 69 |
| | C$_2$H$_6$ | (40) | 2.07 ± 0.10 | 1.82 ± 0.12 | 3.77 ± 0.26 | 3.05 ± 1.22 | 1.00 | 9.2 ± 2.0 |
| | | (50) | 2.21 ± 0.10 | | 4.02 ± 0.28 | 3.25 ± 1.31 | 1.00 | 9.8 ± 2.1 |
| | CH$_3$OH | (40) | 5.33 ± 0.41 | [1.81± 0.09] | 9.65 ± 0.88 | 7.86 ± 1.67 | 2.57 ± 0.23 | 0.24 ± 0.05 |
| | | (50) | 6.31 ± 0.41 | 1.82 ± 0.26 | 11.43 ± 0.95 | 9.31 ± 1.92 | 2.86 ± 0.23 | 0.28 ± 0.06 |
| KL2 | H$_2$O | (40) | 65.6 ± 9.4 | 1.55 ± 0.59 | 117.3 ± 22.2 | 100 | 28.20 ± 4.20 | 287 ± 80 |
| | | (50) | 71.0 ± 9.2 | | 126.9 ± 22.8 | 100 | 24.74 ± 3.72 | 310 ± 84 |
| | HCN | (40) | 0.42±0.04 | 2.12 ± 0.31 | 0.754± 0.111 | 0.64 ± 0.16 | 0.17 ± 0.02 | 1.84 ± 0.46 |
| | | (50) | 0.45±0.05 | | 0.811± 0.134 | 0.64 ± 0.15 | 0.16 ± 0.02 | 1.98 ± 0.52 |
| | C$_2$H$_2$ | (40) | < 0.501 | [1.79± 0.22] | < 0.896 | < 0.76 | < 0.21 | < 2.19 |
| | | (50) | < 0.559 | | < 1.02 | < 0.80 | < 0.20 | < 2.49 |
| | C$_2$H$_6$ | (40) | 2.43 ± 0.18 | 1.68 ± 0.20 | 4.35 ± 0.63 | 3.71 ± 0.57 | 1.00 | 10.6 ± 2.7 |
| | | (50) | 2.87 ± 0.22 | | 5.13 ± 0.74 | 4.04 ± 0.61 | 1.00 | 12.5 ± 3.1 |
| | CH$_3$OH | (50) | 5.49 ± 0.78 | | 9.82 ± 1.85 | 7.73 ± 1.49 | 1.91 ± 0.31 | 24.0 ± 6.6 |
| **2022 Aug 04**, R$_h$ = 2.48 au, Δ = 1.89 au, dΔ/dt = +12.33 km s$^{-1}$** | | | | | | | | |
| KL2 | H$_2$O | (50) | 129±0.28 | 2.01 ± 0.48 | 244 ± 55 | 100 | 71.6 ± 20.2 | 597 ± 182 |
| | HCN | (50) | 63.6±0.85 | 2.24 ± 0.30 | 1.23 ± 0.17 | 0.50 ± 0.10 | 0.35 ± 0.08 | 3.01 ± 0.74 |
| | C$_2$H$_2$ | (50) | < 0.257 | [1.90 ± 0.14] | < 0.49 | < 0.20 | < 0.14 | < 1.20 |
| | CH$_4$ | (50) | 9.10±2.08 | 1.78 ± 0.22 | 42.3 ± 4.1 | 7.08 ± 1.70 | 5.07 ± 1.49 | 42.4 ± 13.2 |
| | C$_2$H$_6$ | (50) | 1.80±0.33 | 1.84 ±0.23 | 3.41 ± 0.68 | 1.40 ± 0.48 | 1.00 | 8.31 ± 2.37 |
| | CH$_3$OH | (50) | 4.90 ± 1.20 | | 9.31 ± 2.39 | 3.81 ± 1.24 | 2.73 ± 0.84 | 22.8 ± 7.4 |
| **2022 Aug 10, R$_h$ = 2.43 au, Δ = 1.93 au, dΔ/dt = 14.24 km s$^{-1}$** | | | | | | | | |
| M2 | CO | $49^{+12}_{-10}$ | 21.2 ± 1.35 | (1.93) | 40.9 ± 8.30 | >16.8 | 8.02 ± 1.80 | 100 |
| | H$_2$O | (49) | < 126 | | < 243 | | < 47.8 | < 597 |
| Lp1 | C$_2$H$_6$ | (49) | 2.65 ± 0.25 | 1.93 ± 0.12 | 5.09 ± 0.49 | > 1.04 | 1.00 | 12.47 ± 2.80 |
| | CH$_4$ | (49) | 6.34 ± 1.58 | | 12.2 ± 3.05 | > 5.02 | 2.40 ± 0.64 | 29.9 ± 9.6 |
| | OH* | (49) | < 236 | [1.93± 0.12] | < 450 | | < 89.1 | < 1100 |
| | CH$_3$OH | (49) | 8.13 ± 1.18 | | 15.7 ± 2.3 | > 6.43 | 3.07 ± 0.53 | 38.3 ± 9.6 |
| **2022 Aug 19, R$_h$ = 2.36 au, Δ = 2.01 au, dΔ/dt = 16.57 km s$^{-1}$** | | | | | | | | |
| Lp1 | $C_2H_6$ | $52^{+7}_{-6}$ | 2.41 ± 0.28 | 2.18 ± 0.23 | 5.25 ± 0.67 | 1.44 ± 0.32 | 1.00 | 8.24 ± 1.25 |
| | $CH_4$ | (52) | 7.30 ± 1.04 | 2.14 ± 0.13 | 15.7 ± 2.3 | 4.29 ± 1.0 | 3.03 ± 0.56 | 24.6 ± 4.1 |
| | $CH_3OH$ | (52) | 7.87 ± 0.85 | [2.15 ± 0.11] | 17.1 ± 1.9 | 4.69 ± 1.0 | 3.27 ± 0.52 | 26.9 ± 3.7 |
| | OH* | (52) | < 148 | | < 323 | >16.2 | < 61.5 | < 507 |
| **2022 Aug 20, $R_h$ = 2.35 au, Δ = 2.02 au, dΔ/dt = 16.76 km s$^{-1}$** | | | | | | | | |
| L-cust | $H_2O$ | (53) | 150 ± 27 | 2.44 ± 0.26 | 365 ± 66 | 100 | 69.6 ± 15.4 | 573 ± 114 |
| | HCN | (53) | 0.72 ± 0.06 | [2.44 ± 0.26] | 1.75 ± 0.16 | 0.48 ± 0.10 | 0.34 ± 0.05 | 2.77 ± 0.34 |
| | $C_2H_2$ | (53) | 0.81 ± 0.14 | | 1.97 ± 0.34 | 0.54 ± 0.13 | 0.38 ± 0.08 | 3.09 ± 0.59 |
| | $NH_3$ | (53) | < 2.56 | | < 6.24 | < 1.71 | < 1.19 | < 9.79 |
| **2022 Aug 21, $R_h$ = 2.35 au, Δ = 2.03 au, dΔ/dt = 16.92 km s$^{-1}$** | | | | | | | | |
| M2 | CO | $53^{+5}_{-4}$ | 24.0 ± 2.0 | 2.65 ± 0.15 | 63.7 ± 5.7 | 17.5 ± 3.5 | 12.1 ± 1.9 | 100 |
| | $H_2O$ | (53) | 189 ± 38 | [2.65 ± 0.15] | < 300 | | < 57.2 | < 471 |
| | CN | (53) | < 0.98 | | < 2.58 | < 0.71 | < 0.49 | < 4.06 |
| | OCS | (53) | < 0.44 | | < 1.2 | < 0.32 | < 0.22 | < 1.84 |

**Notes.**

(a) Rotational temperature determined using 8-15 lines for CO, 4 lines for $CH_4$, and 16-48 lines for $C_2H_6$. Values in parentheses are assumed.

(b) Nucleus centered production rate ($Q_{nc}$).

(c) Growth factor. Values in brackets are assumed.

(d) Global production rate.

(e) Molecular abundance with respect to $H_2O$. *For Aug 19 and 21, $Q(H_2O)$ from Aug 20 was assumed.*

(f) Molecular abundance with respect to $C_2H_6$. *For Aug 20 and 21, $Q(C_2H_6)$ from Aug 19 was assumed.*

(g) Molecular abundance with respect to CO. *For Aug 19 and 20, $Q(CO)$ from Aug 21 was assumed; for July 28 and Aug 4, $Q(CO)$ from Aug 10 was assumed.*

\*\*NIRSPEC dates. Both $C_2H_6$ & $CH_3OH$ are reported for July 28 (from both KL1 & KL2 settings) and Aug 04 (from KL2 only). $CH_3OH$ in KL2 on the 2 dates was determined using an emission model in the 2920 cm$^{-1}$ region (Bonev et al., in prep).

## 4. DISCUSSION

### 4.1 Heliocentric dependence of Production Rates in C/2017 K2 (PanSTARRS)

We investigated pre-perihelion volatile activity in comet K2 between $R_h$ = 3.15 and 2.35 au. Only the hypervolatiles CO, $CH_4$, and $C_2H_6$ were detected beyond ~2.8 au, however, K2 was sufficiently bright and active that $H_2O$, HCN, $C_2H_2$, and $CH_3OH$ were subsequently also detected

(or were constrained to meaningful levels) at smaller $R_h$. The production rates of CO, $CH_4$, and $C_2H_6$ were relatively constant (within uncertainty) beyond ~2.6 au. The overall gas production increased at smaller $R_h$ (owing to increased solar flux) as expected (see Figure 3), but with hypervolatile production rates remaining relatively constant (albeit with a modest increase in Q($CH_4$) between 2.40 and 2.36 au). The CO production rate showed a moderate increase between 3.15 and 2.35 au inbound, with power-law slope $R_h^{-1.8}$, which is generally consistent with the slope measured for Hale-Bopp at $R_h$ ~ 6.7- 0.9 au pre-perihelion (Biver et al. 2002). Combi et al. (2023) reported post-perihelion $H_2O$ production rates for K2, using the Solar Wind ANisotropies (SWAN) instrument onboard the Solar and Heliospheric Observer (SOHO) satellite. The values measured by SWAN are about a factor of two or more higher than our pre-perihelion values for K2 at similar $R_h$. Cometary gas production rates are often asymmetric with respect to perihelion, but even absent this, higher Q($H_2O$) measured by SWAN might be expected given its larger field of view compared with ground-based IR spectrographs, in particular, if additional $H_2O$ was released from icy grains in regions of the coma not encompassed by the IR slit. As an example, this was inferred for comet C/2012 S1 (ISON) over a similar range in pre-perihelion $R_h$ (Combi et al. 2014; DiSanti et al. 2016, § 4.1.2).

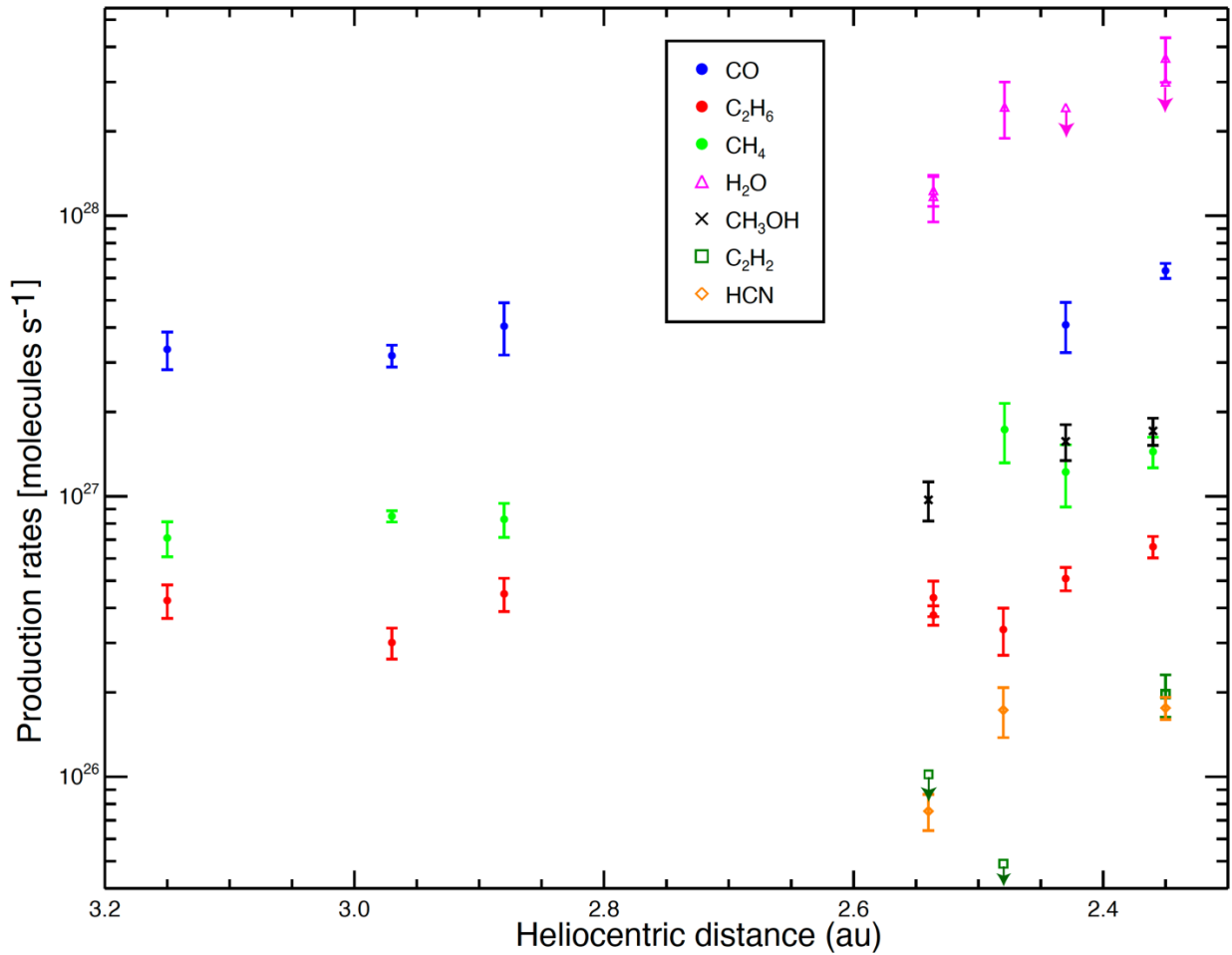

Figure 3. Production rates of CO (blue), $C_2H_6$ (red), $CH_4$ (green), $H_2O$ (magenta), $CH_3OH$ (orange), $C_2H_2$ (olive), and HCN (turquoise) in K2 as a function of heliocentric distance.

### 4.2 Comparison with other OCCs within $R_h$ = 2 au

We also compared the molecular abundances of volatile species, measured with respect to $H_2O$ (as well as to $C_2H_6$) in K2 (from Table 2), to their mean values among OCCs within $R_h$ = 2 au measured at high spectral resolution between $\lambda \sim 2.8 – 5.0$ μm [Fig. 4. (a) and (b)]. Once water (the least volatile ice measured in comets) is fully activated, the abundances of the volatile species relative to $H_2O$ are expected to reflect the actual composition of ices in the nucleus. The molecular abundances in K2 relative to $H_2O$ changed significantly with $R_h$ between 2.54 and 2.35 au with the level of enrichment dropping as the comet approached the Sun, which would be consistent with a scenario in which water was beginning to activate over this heliocentric range. Thus, because $H_2O$ production in K2 was possibly not yet fully activated at the heliocentric distances of our measurements, we compared the molecular abundances of volatile species in K2 ($CH_4$, $CH_3OH$, CO, $C_2H_2$ and HCN) with respect to $H_2O$ (as well as to $C_2H_6$) in K2 to their mean values among OCCs that were observed within $R_h$ = 2 au at high spectral resolution between $\lambda \sim 2.8 - 5.0$ μm [Table 3, Fig. 4. (a) and (b)]. Table 3 clearly shows that the overall abundances relative to water were significantly higher than their mean values from OCCs, with the enhancements being most pronounced for the hypervolatiles CO and $CH_4$. We note that with iSHELL these hypervolatiles were acquired using different instrument settings from that of $H_2O$ (water was not detected in the M2 setting, even at $R_h$ = 2.35 au; see Table 2). Thus, the effect of rotation over the short time intervals between observations in different settings, as for example different surface regions of the nucleus become illuminated (and activated), cannot be ruled out. On the other hand, it is important to note that the abundances (including those of the hypervolatiles) relative to $C_2H_6$ were generally within the range of values from OCCs measured to date within $R_h \sim 2$ au or even slightly depleted (as for $CH_3OH$). The enriched abundances of volatile species with respect to $H_2O$ suggest that either $H_2O$ was not fully activated even at 2.35 au and/or that K2 was unusually enriched in these volatiles, as was the case for comets C/2001 A2 (Gibb et al. 2007; Magee-Sauer et al. 2008), or that a significant amount of $H_2O$ was released by icy grains outside the slit (e.g., Dello Russo et al. 2008). The mixing ratios of hypervolatiles as well as less volatile ices such as HCN decreased with decreasing $R_h$ relative to $H_2O$, consistent with the former possibility. Even though K2 presented a relatively rare opportunity to conduct a detailed IR study of cometary activity at larger Rh, this discussion emphasizes the need to measure additional comets within the critical heliocentric distance range $R_h \sim$2-3 au (or larger).

It is worth noting the mixing ratios for $C_2H_6$, $CH_3OH$, and HCN relative $H_2O$ at 2.54 au, 2.48 au and 2.36 au (see Table 3). The abundances of these volatiles decreased between 2.54 and 2.48 au, and then remained more nearly constant from 2.48 to 2.36 au. We note that $H_2O$ was co-measured within the same instrument settings used for the (Keck/NIRSPEC) measurements at $R_h$ = 2.54 and 2.48 au. On the other hand, for the $C_2H_6$ and $CH_3OH$ abundances relative to $H_2O$ at $R_h$ = 2.36 au, the $H_2O$ measurement was obtained a day later (at 2.35 au), so (as mentioned previously) the possibility of nucleus rotation influencing the abundances of these volatiles relative to $H_2O$ (or overall diurnal changes in gas production, regardless of the cause) cannot be ruled out.

**Table 3. Summary of volatile abundances in K2 and comparisons with OCCs within 2 au**

| Molecule | $R_h$ (au) | $Q(X)/Q(C_2H_6)$ [a] | | $Q(X)/Q(H_2O)$ [%] [a] | |
|---|---|---|---|---|---|
| | | K2 | OCCs [b] | K2 | OCCs [b] |
| $CH_4$ | 3.15 | 1.67 ± 0.32 | 1.42 ± 0.11 | > 1.53 | 0.77 ± 0.08 |
| | 2.97 | 2.82 ± 0.37 | | > 3.46 | |
| | 2.88 | 1.92 ± 0.37 | | > 2.54 | |
| | 2.48 | 5.16 ± 1.58 | | 7.08 ± 1.70 | |
| | 2.43 | 2.40 ± 0.64 | | > 5.02 | |
| | 2.36 | 2.98 ± 0.55 | | 4.29 ± 1.0 | |
| $C_2H_6$ | 3.15 | 1.00 | 1.00 | > 0.92 | 0.59 ± 0.08 |
| | 2.97 | | | > 1.23 | |
| | 2.88 | | | > 1.34 | |
| | 2.54 | | | 3.90 ± 0.55 | |
| | 2.48 | | | 1.40 ± 0.48 | |
| | 2.36 | | | 1.44 ± 0.32 | |
| CO | 3.15 | 8.10 ± 1.67 | 8.82 ± 1.99 | > 7.62 | 4.97 ± 1.24 |

|      | 2.97 | 11.1 ± 1.8 |            | > 13.8      |             |
|------|------|------------|------------|-------------|-------------|
|      | 2.88 | 8.86 ± 2.22 |           | >12.0       |             |
|      | 2.43 | 8.02 ± 1.80 |           | > 16.8      |             |
|      | 2.35 | 12.1 ± 1.9 |            | 17.5 ± 3.5  |             |
| $CH_3OH$ | 2.54 | 2.52 ± 0.19 | 3.92 ± 0.35 | 8.32 ± 1.18 | 2.11 ± 0.19 |
|      | 2.48 | 2.73 ± 0.84 |            | 3.81 ± 1.24 |             |
|      | 2.43 | 3.07 ± 0.53 |            | > 6.43      |             |
|      | 2.36 | 3.27 ± 0.52 |            | 4.69 ± 1.00 |             |
| HCN  | 2.54 | 0.16 ± 0.02 | 0.38 ± 0.04 | 0.64 ± 0.16 | 0.20 ± 0.02 |
|      | 2.48 | 0.35 ± 0.08 |            | 0.50 ± 0.10 |             |
|      | 2.35 | 0.34 ± 0.05 |            | 0.48 ± 0.10 |             |
| $C_2H_2$ | 2.54 | < 0.20 | 0.28 ± 0.05 | < 0.80 | 0.14 ± 0.02 |
|      | 2.48 | < 0.14 |            | < 0.20 |             |
|      | 2.35 | 0.38 ± 0.08 |           | 0.54 ± 0.13 |             |

Notes: The $C_2H_6/H_2O$, $CH_3OH/H_2O$, and $CH_3OH/C_2H_6$ abundance values indicated at 2.54 au are the weighted mean of the abundances for KL1 and KL2 indicated in Table 2.

[a] Mean abundances relative to simultaneously (or contemporaneously) measured $C_2H_6$ and $H_2O$ in comet K2.

[b] Mean abundances among OCCs, taken from Dello Russo et al. (2016) and results for comets observed (published) subsequently: C/2020 S3 (Ejeta et al. 2024), C/2017 E4 (Faggi et al. 2018), C/2013 V5 (DiSanti et al. 2018), C/2021 A1 (Faggi et al. 2023; Lippi et al. 2023), C/2002 T7 and C/2015 ER61 (Saki et al. 2021), C/2018 Y1 (DiSanti et al. 2021), C/2014 Q2 (Dello Russo et al. 2022), and C/2012 K1 (Roth et al. 2017).

### 4.3 Comparison with other OCCs beyond $R_h$ = 2 AU

To put K2 in context, we also compare its volatile abundances to those of other comets observed beyond ~2.0 au [Fig. 4 (c), (d), (e) and (f)]. The $CH_4$ abundance relative to CO in K2 was constant (within uncertainties) between $R_h$ = 3.15 and 2.35 au (Fig. 4(c)); however, it was significantly higher than observed for C/2006 W3 (Christensen) and C/2016 R2 (Pan-STARRS) at similar $R_h$ (Bonev et al. 2017; Bockéele-Morvan et al. 2010; McKay et al. 2019; Wierzchos & Womack 2018), comets for which CO dominated the outgassing activity [these are also shown in Fig. 4 (c)]. In addition, as shown in Fig.4 (e), the CO abundance relative to $C_2H_6$ in K2 was lower than that for comets C/1995 O1 (Hale-Bopp) and C/2006 W3 at similar $R_h$ (see Bonev et al. 2017; Bockéele-Morvan et al. 2010; DiSanti et al. 2001; Dello Russo et al. 2001). Again, this is due to these comets being significantly CO-rich compared to K2. On the other hand, Outsubo et al. (2012) observed a number of comets spanning $R_h$ = 2.42 - 3.29 au and obtained only upper limits for CO production that were of order ~$10^{26}$ molecules/s, giving only 3-$\sigma$ upper limits (ranging from ~4-10 %) for the CO abundance (relative to $H_2O$) for some of these comets [Fig. 4 (f)]. Thus, more measurements of CO and $C_2H_6$ in comets beyond ~2 au are required to draw more definitive conclusions. We also note that even though comet 17P/Holmes had an outburst few days prior to its post-perihelion observations at heliocentric distances 2.45 – 2.47 au, the measurements show enhanced abundances for volatiles $C_2H_6$, $CH_3OH$, HCN, and $C_2H_2$ relative to $H_2O$ (Dello Russo et al. 2008). Similarly, post-perihelion observations of comet C/2010 G2 (Hill) at $R_h$ ~2.5au, about a week following its outburst, show higher abundances of volatiles CO, $CH_4$, $C_2H_6$, HCN, and $CH_3OH$ relative to $H_2O$ (Kawakita et al. 2014).

Apart from $R_h$-dependence, the overall production rates depend on the radius of the nucleus, the active fractional surface area, whether the comet is observed pre- or post-perihelion (e.g., DiSanti et al. 2014, Roth et al. 2018), and the dust-to-gas ratio. Yang et al. (2021) measured a 3.6$\sigma$ CO detection at $R_h$ = 6.72 au and dust/gas ratio > 1 in comet K2. Using imaging observations, Jewitt et al. (2017) determined the radius of the nucleus of K2 to be ≤ 9 km, smaller by a at least a factor of ~2-5 compared to Hale-Bopp, which was estimated to have a radius of 30 ±10 km (Fernández 2002). Even though the size of C/2016 R2 is unknown, Wierzchos et al. (2018) suggested its radius to be ≤ ~15 km assuming its CO production rate is proportional to the surface area of its nucleus. Even though there were no accurate size measurements for comets C/2006 W3 and C/2016 R2, it is possible that the significantly lower CO production rates we measured for K2 compared to those measured for these comets at similar $R_h$ could be due to, among other factors, the dusty nature of K2 and/or its smaller inferred size.

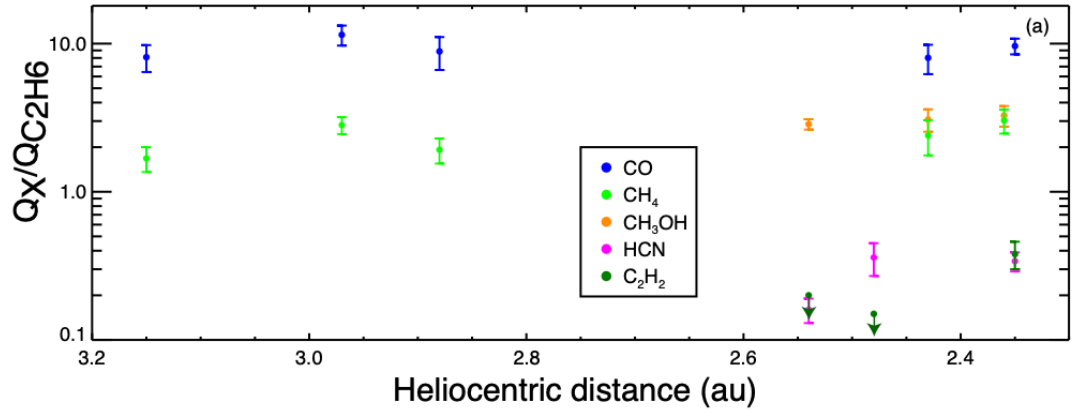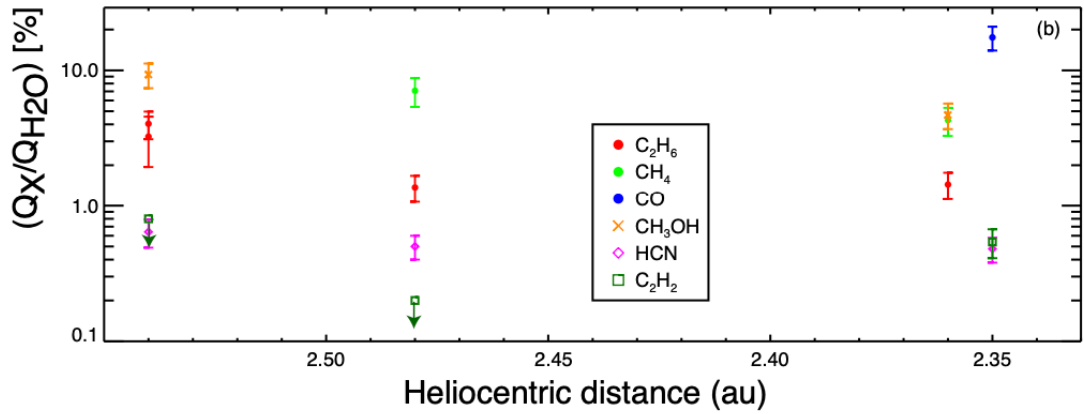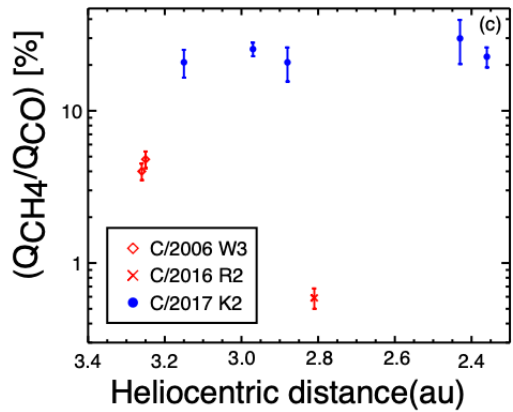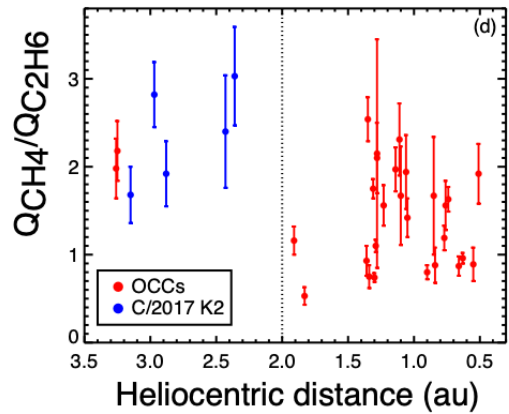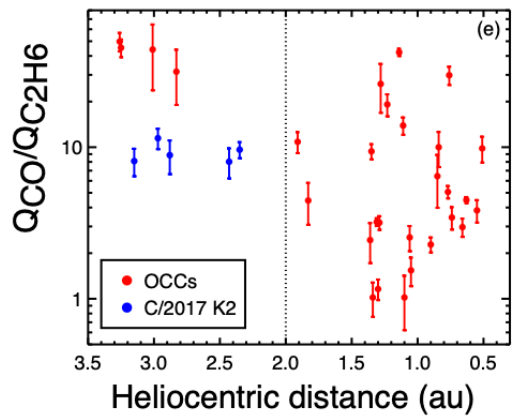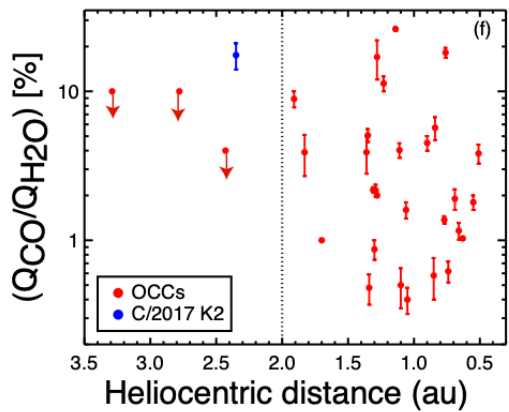

Figure 4. The abundances of volatiles in comet K2 [panels (a) and (b)], abundances in K2 and OCCs measured at comparable heliocentric distances [panel (c)], and comparison of abundances of volatiles in K2 in comparison with OCCs measured to date in the infrared wavelength [panels (d), (e) and (f)]. The vertical dotted lines indicate measurements within 2 au, where most measurements are performed.

## 5. SUMMARY

We obtained pre-perihelion observations of comet K2 covering a heliocentric distance range $R_h \sim 3.15 - 2.35$ au, to measure the evolution of hypervolatiles CO, $CH_4$, and $C_2H_6$, and to test the activity, especially the onset of $H_2O$ sublimation at detectable levels, and to track the development of other volatiles ($C_2H_2$, $NH_3$, HCN, $CH_3OH$, CN, and OCS).

Abundances of volatile species with respect to simultaneously or contemporaneously observed $H_2O$ in K2 were significantly higher than the mean values from OCCs measured to date, most of which were measured within 2.0 au from the Sun. On the other hand, the abundances of these volatiles relative to $C_2H_6$ remained more consistent with $R_h$ and were generally also consistent with their mean values from OCCs (see Table 3). This could imply that comet K2 was either enriched in these (normally "trace") volatiles (or was depleted in water), or that $H_2O$ was not yet fully activated beyond 2.0 au. The increase in water production rate near 2.40 au that was not seen to the same degree in other molecules suggests the latter possibility. Clearly, more observations of comets are needed over the critical range $R_h \sim 2 - 3$ au, to more conclusively establish the heliocentric distance at which (or the range of $R_h$ over which) cometary $H_2O$ becomes fully activated, as well as to test the behavior of ice sublimation in comets beyond 2 au.

**Acknowledgments**

Data for this study were obtained at the NASA Infrared Telescope Facility (IRTF), operated by the University of Hawai′i under contract NNH14CK55B with NASA, and the W.M.

Keck Observatory, operated as a scientific partnership among the California Institute of Technology, the University of California, and NASA. The W.M. Keck Observatory was made possible by the generous financial support of the W. M. Keck Foundation. The authors gratefully acknowledge the support of the IRTF and Keck 2 staff for their help in acquiring the K2 data. E.G. and C.E. acknowledge support from NSF under award 2009910. M.D. and N.X.R. acknowledge support through the NASA Solar System Observations Program under award 22-SSO22_0013. B.P.B acknowledges the support of NSF grant under the award: AST-2009398.

**References**

A'Hearn, M., Feaga, L. M., Keller, H. U., et al. 2012, ApJ, 758, 29

https://doi.org/10.1088/0004-637X/758/1/29

Biver, N., Bockelée-Morvan, D., Crovisier, J., et al. 2006, A&A, 449, 1255

https://doi.org/10.1051/0004-6361:20053849

Biver, N., Bockelée-Morvan, D., Pierre, C., et al. 2002, EM&P, 90, 5

https://doi.org/10.1023/A:1021599915018

Bockelée-Morvan, D., Hartogh, P., Crovisier, J., et al. 2010, A&A, 518, L149

https://doi.org/10.1051/0004-6361/20101465

Bockelée-Morvan, D, Crovisier, J., Mumma, M. J., Weaver, H.A., & Binzel, R. P., 2004, in Comets II, Weaver, H.A., Festou, M.C., and Keller, H. U. (eds.), University of Arizona Press, Tucson,745 pp., p. 391- 424

Bonev, B. P., Dello Russo, N., DiSanti, M. A., et al. 2021, PSJ, 2, 45

https://doi.org/10.3847/PSJ/abd03c

Bonev, B.P., Villanueva, G. L., DiSanti, M. A., et al. 2017, AJ, 153, 241

https://doi.org/10.3847/1538-3881/aa64dd

Bonev, B. P., Mumma, M. J., Gibb, E. L., et al. 2009, ApJ, 699, 1563

https://doi.org/10.1088/0004-637X/699/2/1563

Bonev, B. P., Mumma, M. J., Villanueva, G. L., et al. 2007, ASJ, 661, L97

https://doi.org/10.1086/518419

Bonev, B. P., and Mumma, M. J. 2006, ApJ, 653, 788

https://doi.org/10.1086/508450

Bonev 2005, Ph.D. Thesis, the University of Toledo

https://ui.adsabs.harvard.edu/abs/2005PhDT.......258B

Cambianica, P., Munaretto, G., Cremonese, G., et al. 2023, A&A, 674, L14

https://doi.org/10.1051/0004-6361/20224555

Combi, M, Mäkinen, T., Bertaux, J-L., Quemerais, E., Ferron, S., 2023, DPS, 55, 322.04

https://ui.adsabs.harvard.edu/abs/2023DPS....5532204C

Combi, M. R., Fougere, N., Mäkinen, T. T., et al. 2014, APJL, 788:L7

https://doi.org/10.1088/2041-8205/788/1/L7

Combi, M. R., Harris, W.M., Smyth, W.H., 2004, Comets II, Festou, M.C., Keller, H. U, and Weaver, H. A. (eds.), University of Arizona Press, Tucson, 745 pp., p.523-552

Crovisier, J., Encrenaz, T., Lyle, S., and Bonnet, R. M., 2000, Cambridge Univ. press, Cambridge, U.K.

Dello Russo, N., Vervack Jr., R. J., Kawakita, H., et al. 2022, PSJ, 3, 6

https://doi.org/10.3847/PSJ/ac323c

Dello Russo, N., Kawakita, H., Bonev, B. P., et al. 2020, Icar, 335, 113411

https://doi.org/10.1016/j.icarus.2019.113411

Dello Russo, N., Kawakita, H., Vervack Jr., R. J., & Weaver, H. A. 2016, Icar, 278, 301

https://doi.org/10.1016/j.icarus.2016.05.039

Dello Russo, N., Vervack Jr., R. J., Weaver, H. A., & Lisse, C. M. 2009, Icar, 200, 271

https://doi.org/10.1016/j.icarus.2008.11.008

Dello Russo, N., Vervack Jr., R. J., and Weaver, H. A., et al. 2008, ApJ, 680: 793-802

https://doi.org/10.1086/587459

DiSanti, M. A., Bonev, B. P., Dello Russo, N., et al. 2021, PSJ, 2, 225

https://doi.org/10.3847/PSJ/ac07ae

DiSanti, Michael A., Bonev, B. P., Gibb, E., L., et al. 2018, AJ, 156, 258

https://doi.org/10.3847/1538-3881/aade87

DiSanti, M. A., Bonev, B. P., Dello Russo, N., et al. 2017, AJ, 154, 246

https://doi.org/10.3847/1538-3881/aa8639

DiSanti, M. A., Bonev, B. P., Gibb, E. L., et al. 2016, APJ, 820, 34

https://doi.org/10.3847/0004-637X/820/1/34

DiSanti, M. A., Villanueva, G. L., Paganini, L., et al. 2014, Icar, 228, 167

https://doi.org/10.1016/j.icarus.2013.09.001

DiSanti, M. A., Bonev B. P., Villanueva, G. L., & Mumma, M. J. 2013, ApJ, 763, 1

https://doi.org/10.1088/0004-637X/763/1/1

DiSanti, M. A., Bonev, B. P., Magee-Sauer, K., et al. 2006, ApJ, 650, 470

https://doi.org/10.1086/507118

DiSanti, M. A., Mumma, M. J., Dello Russo, N., & Magee-Sauer, K. 2001, Icar, 153, 361

https://doi.org/10.1006/icar.2001.6695

Ejeta, C., Gibb, E., Roth, N., et al. 2024, AJ, 167, 32

https://doi.org/10.3847/1538-3881/ad0e02

Faggi, S., Lippi, M., Mumma, M. J., Villanueva, G. L., 2023, PSJ, 4,8

https://doi.org/10.3847/PSJ/aca64c

Faggi, S., Villanueva, G. L., Mumma, M. J., Paganini, L., 2018, AJ, 156, 68

https://doi.org/10.3847/1538-3881/aace01

Fernández, Y. R., 2002, EM&P, 89,3-25

https://doi.org/10.1023/A:1021545031431

Fougere, N., Combi, M. R., Tenishev, V., et al. 2012, Icar, 221, 174

https://dx.doi.org/10.1016/j.icarus.2012.07.019

Gibb, E. L., Bonev, B. P., Villanueva, G., et al. 2012, ApJ, 750, 102

https://doi.org/10.1088/0004-637X/750/2/102

Gibb, L. E., Disanti, M. A., Magee-Sauer, K., et al. 2007, Icar, 188, 224

https://doi.org/10.1016/j.icarus.2006.11.009

Gibb, E. L., Mumma, M. J., Dello Russo, N., DiSanti, M. A., & Magee-Sauer, K. 2003, Icar, 165, 391

https://doi.org/10.1016/S0019-1035(03)00201-X

Guilbert-Lepoutre, A., 2012, AJ, 144, 97

https://doi.org/10.1088/0004-6256/144/4/97

Hui, M-T., Jewitt, D., Clark, D., 2018, AJ, 155, 25

https://doi.org/10.3847/1538-3881/aa9be1

Jehin, E., Vander Donckt, M., Hmiddouch, S., Manfroid, J. and Hutsemekers, H. 2022, ATel, 15591,1

https://ui.adsabs.harvard.edu/abs/2022ATel15591....1J

Jewitt, D., Hui, M-T., Mutchler, M., et al. 2017, ApJL, 847, L19

https://doi.org/10.3847/2041-8213/aa88b4

Jewitt, D., 2009, AJ, 137, 4296

https://doi.org/10.1088/0004-6256/137/5/4296

Kawakita, H., Dello Russo, N., Vervack Jr., R., et al. 2014, ApJ, 788, 110

https://doi.org/10.1088/0004-637X/788/2/110

Kawakita, H. & Mumma, M. J. 2011, APJ, 727, 91

https://doi.org/10.1088/0004-637X/727/2/91

Lippi, M., Vander Donckt. M., Faggi, S., et al. 20203, A&A, 676, A105

https://doi.org/10.1051/0004-6361/202346775

Magee-Sauer, K., Mumma, M. J., DiSanti, M. A., et al. 2008, Icar, 194, 347

https://doi.org/10.1016/j.icarus.2007.10.006

Martin, E. C., Fitzgerald, M. P., McLean, I. S., et al. 2018, Proc. SPIE, 10702, 107020A

https://doi.org/10.1117/12.2312266

Martin, E. C., Fitzgerald, M. P., McLean, I., Kress, E., Wang, E. 2016, Proc. SPIE, 9908, 99082R

https://doi.org/10.1117/12.2233767

McKay, A. J., DiSanti, M. A., Kelley, M. S. P., et al. 2019, AJ, 158, 128

https://doi.org/10.3847/1538-3881/ab32e4

Meech, K. J., Kleyna, J. T., Hainaut, O., et al. 2017, ApJL, 849, L8

https://doi.org/10.3847/2041-8213/aa921f

Meech, K. J., Svoren, J., 2004, Comets II, Festous, M. C., Keller, H. U., and Weaver, H. A. (eds),

University of Arizona Press, Tucson, 745pp., p.317-335

Mumma, M. J., & Charnley, S. B., 2011, Annu. Rev. Astron. Astrophys, 49, 471

https://doi.org/10.1146/annurev-astro-081309-130811

Ootsubo, T., Kawakita, H., Hamada, S., et al. 2012, ApJ, 752, 15

https://doi.org/10.1088/0004-637X/752/1/15

Paganini, L., DiSanti, M. A., Mumma, M. J., et al. 2014, AJ, 147, 15

https://doi.org/10.1088/0004-6256/147/1/15

Paganini, L., Mumma, M. J., Villanueva, G. L., et al. 2012, ApJL, 748, L13

https://doi.org/10.1088/2041-8205/756/2/L42

Prialnik, D., 1992, ApJ, 388, 196

https://doi.org/10.1086/171143

Prialnik, D., Bar-Nun, A., 1992, A&A, 258, L9

https://ui.adsabs.harvard.edu/abs/1992A&A...258L...9P

Radeva, Y. L., Mumma, M. J., Bonev, B. P., et al. 2010, Icar, 206, 764

https://doi.org/10.1016/j.icarus.2009.09.014

Rayner, J., Tokunaga, A., Jaffe, D., et al. 2022, PASP, 134:015002

https://doi.org/10.1088/1538-3873/ac3cb4

Roth, N.X., Gibb, E. L., Bonev, B.P., et al. 2018, AJ, 156:251

https://doi.org/10.3847/1538-3881/aae0f7

Roth, N. X., Gibb, E. L., Bonev, B. P., et al. 2017, AJ, 153, 168

https://doi.org/10.3847/1538-3881/aa5d18

Saki, M., Gibb, E. L., Bonev, B.P. et al. 2021, AJ, 162, 145

https://doi.org/10.3847/1538-3881/abfcbd

Schleicher, D. G., Millis, R. L., Brich, P. V., 1998, Icar, 132, 397

https://doi.org/10.1006/icar.1997.5902

Villanueva, G. L., Smith, M. D., Protopapa, S., Faggi, S., Mandell, A. M., 2018, JQSRT, 217, 86

https://doi.org/10.1016/j.jqsrt.2018.05.023

Villanueva, G. L., Magee-Sauer, K., & Mumma, M. J. 2013, JQSRT, 129, 158

https://doi.org/10.1016/j.jqsrt.2013.06.010

Villanueva, G. L., Mumma, M. J., Bonev B. P., et al. 2012a, JQSRT, 113, 202

https://doi.org/10.1016/j.jqsrt.2011.11.001

Villanueva, G. L., Disanti, M. A., Mumma, M. J., & Xu, L. -H. 2012b, ApJ, 747, 37

https://doi.org/10.1088/0004-637X/747/1/37

Villanueva, G. L., Mumma, M. J., & Magee-Sauer, K. 2011a, J. Geophys. Res., 116, E08012

https://doi.org/10.1029/2010JE003794

Villanueva, G. L., Mumma, M.J., DiSanti, M.A., et al. 2011b, Icar, 2016, 227

https://doi.org/10.1016/j.icarus.2011.08.024

Wainscoat, R.J., Wells, L., Micheli, M., Sato, H. 2017, CBET, 4393, 1

https://ui.adsabs.harvard.edu/abs/2017CBET.4393....1W

Wierzchos, K., Womack, M. 2018, AJ, 156, 34

https://doi.org/10.3847/1538-3881/aac6bc

Willacy, K., Turner, N., Bonev, B., et al. 2022, ApJ, 931, 164

https://doi.org/10.3847/1538-4357/ac67e3

Womack, M., Curtis, O., Rabson, D. A., et al. 2021, PSJ, 2, 17

https://doi.org/10.3847/PSJ/abd32c

Xie, X., and Mumma, M. J., 1996, ApJ, 464, 457

https://doi.org/10.1086/177336

Yang, B., Jewitt, D., Zhao, Y., et al. 2021, ApJL, 914, L17

https://doi.org/10.3847/2041-8213/ac03b7